\documentclass[aps,prb,twocolumn,nofootinbib]{revtex4}
\usepackage{amsmath} 
\usepackage{amsfonts} 
\usepackage{graphicx} 
\usepackage{stix}
\usepackage[export]{adjustbox}

\begin{document}
\title{Testing the necessity of complex numbers in traditional quantum theory with quantum computers}
\author{Jarrett L. Lancaster}
\affiliation{Department of Arts \& Sciences, Maine Maritime Academy, 1 Pleasant Street, Castine, ME 04420}
\email{jarrett.lancaster@mma.edu}
\author{Nicholas M. Palladino}
\affiliation{Department of Physics, High Point University, One University Parkway, High Point, NC 27262}

\date{\today}

\begin{abstract}
A recent experiment testing the necessity of complex numbers in the standard formulation of quantum theory is recreated using IBM quantum computers. To motivate the experiment, we present a basic construction for real-valued quantum theory. The real-valued description is shown to predict correlations identical to those of complex-valued quantum mechanics for two types of Bell tests based on the Clauser-Horne-Shimony-Holt (CHSH) inequality. A slight modification to one test, however, results in different predictions for the real- and complex-valued constructions. While noisier devices are incapable of delivering convincing results, it is shown that certain devices possess sufficiently small error rates to falsify real-valued formulations of quantum theory for composite states. The results obtained with quantum computers are consistent with published experiments. This work demonstrates the feasibility of using freely-available quantum devices to explore foundational features of quantum mechanics with minimal technical expertise. Accordingly, this treatment could inspire novel projects for undergraduate students taking a course on quantum mechanics. 
\end{abstract}

\maketitle
\section{Introduction}
\footnote{Accepted by {\it Am. J. Phys.}. After it is published, it will be found at:\\ \href{https://pubs.aip.org/aapt/ajp}{https://pubs.aip.org/aapt/ajp}}This article shows how current quantum hardware can be used to realize small entangled systems with sufficient coherence to perform nontrivial experiments. Cleverly-designed small experiments are ideal settings for probing fundamental aspects of quantum mechanics. Previous work has shown that Bell's theorem~\cite{Bell} and its extensions provide numerous opportunities~\cite{DallaTorre2022,Wang2021,Sadana2022,Santini2022,BrodyTPT} for these noisy devices to confirm experimentally the validity of the traditional framework of quantum mechanics. This article focuses on the important role of complex numbers in the traditional formulation of quantum theory.

In classical physics, complex numbers are often used to perform calculations efficiently. For example, standard treatments of the damped, driven oscillator and electromagnetic waves leverage Euler's identity $\cos \phi  = \mbox{Re}[e^{i\phi}]$ and the linear nature of these systems to streamline certain calculations. The use of complex variables in these classical settings appears only in intermediate steps of calculations, with real input leading to real output. But as the imaginary unit $i$ appears explicitly in the Schr\"{o}dinger equation, the role of complex numbers in quantum mechanics might {\it appear} to be more fundamental than in classical physics. Indeed, convincing justification for the necessity of complex numbers in quantum theory has been presented in this journal by Karam.~\cite{Ricardo} Interestingly, the choice of using complex or real numbers also has important consequences for the structure of the Hilbert spaces which describe multiparticle states.~\cite{Avella2022} 

Recently, a Bell-like test was devised which can distinguish between the different Hilbert space structures required by real- or complex-valued quantum theory.~\cite{SciAm,Renou2021} The standard postulates of quantum mechanics based on Hilbert space~\cite{NielsenChuang} are assumed by construction. Remarkably, {\it any} such real-valued formulation can be shown to result in a lower correlation than complex-valued quantum theory. Realizations of this Bell test using superconducting qubits~\cite{qmtest1} and optical quantum networks\cite{qmtest2} have confirmed that an accurate description of nature which adopts the standard postulates of quantum theory appears to require complex numbers. In this article, we show how one can use freely-available quantum computers to perform the same experiment.

It should be noted that there are real-valued constructions of quantum theory~\cite{McKaguePRL} that reproduce all predictions of complex-valued quantum theory at the expense of violating one or more of the fundamental postulates of traditional quantum theory. In particular, modifying the tensor-product structure of composite states provides a potentially innocuous way to obtain a real-valued description of quantum mechanics which matches the predictions of complex-valued constructions. Hypothetical, alternative constructions (e.g., pilot wave theory, or Bohmian mechanics\cite{Bohm1,Bohm2}) using only real numbers also cannot be ruled out. Moreover, Bell tests based upon the Clauser-Horne-Shimony-Holt (CHSH) inequality~\cite{CHSH} used in this work have potential loopholes that make drawing definitive conclusions difficult, even with carefully designed experiments. Experimental violation of a Bell inequality is typically used as evidence to invalidate local realism. Such loopholes allow for explanation of the violation without abandoning local realism.\cite{Larsson2014}

Only recently have so-called ``loophole-free'' experiments been performed for classic tests of Bell's inequality.\cite{Hensen2015,Giustina2015,Shalm2015} Employing cloud-based quantum computers to perform Bell tests does not allow one to address such loopholes. Potential loopholes still exist in experimental tests which claim to demonstrate the necessity of complex numbers in the traditional formulation of quantum theory, but recent progress has been made in closing several.\cite{Wu2022} Exploring how to close such loopholes is beyond the scope of this work and is likely not possible using cloud-based quantum devices. Acknowledging these important limitations and caveats, we show that IBM's freely-available and easy-to-use devices are capable of obtaining results comparable to those obtained in much more specialized experimental setups.\cite{qmtest1,qmtest2} This work demonstrates the potential value of this new technology for students and nonexperts to explore fundamental aspects of quantum theory with minimal cost and experimental expertise. 

In Sec. II, we sketch a possible real-valued construction of quantum theory for spin-$\frac{1}{2}$ systems and show that the predictions of this formulation are indistinguishable from those of complex-valued quantum mechanics in a Bell test involving the CHSH inequality. A more sophisticated ``triple'' CHSH test highlights potential shortcomings of the real-valued description. A generalization of this triple CHSH inequality yields apparent differences between the real- and complex valued constructions. The experimental scheme\cite{Renou2021} for testing real-valued quantum theory is described in Sec.~\ref{sec:exp}, which also contains details of the implementation using IBM quantum hardware. Results from performing this experiment on several IBM Quantum devices are given in Sec.~\ref{sec:res}. Lastly, Sec.~\ref{sec:conc} contains a discussion of the main results and potential directions for further investigation.

\section{Real-valued quantum mechanics}\label{sec:real}
There are numerous ways in which complex numbers affect the structure of quantum theory.\cite{Ricardo} At a basic level, complex numbers might appear immediately necessary since the imaginary unit $i\equiv \sqrt{-1}$ appears prominently in the most general form of the Schr\"{o}dinger equation,
\begin{eqnarray}
i\hbar \frac{d}{dt}|\psi\rangle & = & \hat{\mathcal{H}}|\psi\rangle.\label{eq:schrod}
\end{eqnarray}
Equation~(\ref{eq:schrod}) governs the time evolution of a quantum state $|\psi\rangle$ given the relevant Hamiltonian operator, $\hat{\mathcal{H}}$. In terms of coordinate-representation wavefunctions, $\psi(x) = \langle x |\psi\rangle$, one could imagine splitting the complex $\psi$ into real and imaginary parts, $\psi(x) = \psi_{R}(x) + i\psi_{I}(x)$, where $\psi_{R,I}(x)$ are real quantities. The Schr\"{o}dinger equation then reduces to two real---but {\it coupled}\cite{Ricardo}---equations. 

To motivate a real-valued framework which turns out to be directly falsifiable, let us summarize the relevant, fundamental postulates~\cite{NielsenChuang} of what we refer to as  ``standard quantum theory.''
\begin{enumerate}
\item A quantum state is represented by a unit vector $|\Psi\rangle$ in a Hilbert space known as the ``state space''---a complex vector space equipped with an inner product $\langle \Psi | \Psi\rangle$.
\item A measurement $\hat{\mathcal{M}}$ is represented by a collection of projection operators $\left\{|m\rangle\langle m|\right\}$ which act on the Hilbert space.
\begin{enumerate}
\item The measurement projections are complete in the sense
\begin{eqnarray}
\sum_{m}|m\rangle\langle m| & = & \hat{I}.
\end{eqnarray}
\item The probability $p(m)$ of obtaining a measurement result labeled by $m$ is given by
\begin{eqnarray}
p(m) & = & \langle \Psi | m\rangle\langle m|\Psi\rangle = |\langle m|\Psi\rangle|^{2}.
\end{eqnarray}
\end{enumerate}
\item For two systems described by quantum states $|\Psi_{1}\rangle$ and $|\Psi_{2}\rangle$, the quantum state of the composite system is given by the tensor product of the individual states, $|\Psi\rangle = |\Psi_{1}\rangle\otimes |\Psi_{2}\rangle$.
\end{enumerate}
There are additional postulates concerning the time evolution of states and additional details surrounding the measurement process. The list presented above contains all the defining features of quantum theory which are needed in this work. 

The first two postulates, while stated quite formally, should be familiar to students of quantum mechanics. The third postulate is typically encountered only when investigating multiparticle systems or exploring the addition of angular momenta. This last postulate is quite delicate, as a general two-particle state cannot be decomposed into a simple product of single-particle states. However, any state in the combined space can be represented by a linear combination of such product states.\cite{Dirac} Even with these caveats in mind, the composition postulate turns out to be a limiting constraint on real-valued representations of quantum theory. Essentially, the tensor product is a simple construction that allows operators to be defined for the composite system such that an operator $\hat{\mathcal{A}}$ acting on $|\Psi_{1}\rangle$ will have trivial action on $|\Psi_{2}\rangle$ in the composite system, and similarly for $\hat{B}$ acting on $|\Psi_{2}\rangle$. Explicitly,
\begin{eqnarray}
\hat{A}\otimes\hat{B} |\Psi_{1}\rangle\otimes|\Psi_{2}\rangle \equiv \left(\hat{A}|\Psi_{1}\rangle\right)\otimes\left(\hat{B}|\Psi_{2}\rangle\right).
\end{eqnarray}

Intuitive justification for this construction can be found by trying to define a total Hamiltonian operator for two non-interacting subsystems. Suppose $\hat{\mathcal{H}}_{i}|\Psi_{i}\rangle = E_{i}|\Psi_{i}\rangle$ for $i=1,2$. In the absence of interactions, the energy of the composite system should be $E_{1}+E_{2}$. A natural first guess for the composite state might be a linear combination of the states $|\Psi_{i}\rangle$. Even if the states exist in the same Hilbert space, a linear combination of states with different energies is generally not an eigenstate of either Hamiltonian. Furthermore, the Hamiltonian cannot be $\hat{\mathcal{H}} = \hat{\mathcal{H}}_{1}+\hat{\mathcal{H}}_{2}$ since these operators might exist in completely different Hilbert spaces. For example, $\hat{\mathcal{H}}_{1}$ could describe the continuous, spatial degrees of freedom of an electron in a Coulomb potential, and $\hat{\mathcal{H}}_{2}$ might be a two-dimensional spin Hamiltonian. Letting $\hat{I}_{d}$ denote the $d$-dimensional identity operator, the proper construction $|\Psi\rangle = |\Psi_{1}\rangle\otimes|\Psi_{2}\rangle$ and $\hat{\mathcal{H}} = \hat{\mathcal{H}}_{1}\otimes \hat{I}_{d_{2}} + \hat{I}_{d_{1}}\otimes\hat{\mathcal{H}}_{2}$ ensures
\begin{eqnarray}
\hat{\mathcal{H}}|\Psi\rangle & = & \left[\hat{\mathcal{H}}_{1}\otimes\hat{I}_{d_{2}}+\hat{I}_{d_{1}}\otimes\hat{\mathcal{H}}_{2}\right]|\Psi_{1}\rangle\otimes|\Psi_{2}\rangle,\nonumber\\
& = & \left(\hat{\mathcal{H}}_{1}|\Psi_{1}\rangle\right)\otimes |\Psi_{2}\rangle + |\Psi_{1}\rangle\otimes\left(\hat{\mathcal{H}}_{2}|\Psi_{2}\rangle\right),\nonumber\\
& = & (E_{1}+E_{2})|\Psi\rangle.
\end{eqnarray}

\subsection{A real-valued representation of quantum theory}
In this section, we modify the first postulate to involve only real coefficients in the underlying quantum state space and explore some basic consequences of this restriction. A formal method for developing a real-valued description of quantum systems was initiated by Stueckelberg.\cite{Stueckelberg} The core idea of Stueckelberg's approach is to introduce a real-valued operator playing the role of the imaginary unit which squares to the negative of the identity operator and commutes with all observables.\cite{Aleksandrova} In order to isolate the difficulties in using a real-valued description, let us specialize to the case of spin-$\frac{1}{2}$ systems. Taking the standard basis defined by $\hat{S}^{z}\left|\uparrow\right\rangle =+\frac{\hbar}{2}\left|\uparrow\right\rangle$ and $\hat{S}^{z}\left|\downarrow\right\rangle =-\frac{\hbar}{2}\left|\downarrow\right\rangle$, the general state $|\psi\rangle = \alpha\left|\uparrow\right\rangle + \beta \left|\downarrow\right\rangle$ can be represented by a two-dimensional vector with complex components
\begin{eqnarray}
|\psi\rangle & \dot{=} & \left(\begin{array}{c} \alpha\\ \beta\end{array}\right).
\end{eqnarray}
Folllowing McIntyre~\cite{McIntyre} and Sakurai~\cite{Sakurai}, we use the symbol $\dot{=}$ to indicate ``represented by'' when giving explicit vector and matrix representations states and operators. The coefficients $\alpha$, $\beta$ are subject to the constraint $|\alpha|^{2} + |\beta|^{2} = 1$. Additionally, the state $e^{i\delta}|\psi\rangle$ is physically indistinguishable from the state $|\psi\rangle$ for any real phase $\delta$ (see Postulate 2(b)). These two constraints mean any state may be represented uniquely in terms of two real numbers. For example, the Bloch sphere parameterization gives
\begin{eqnarray}
|\psi\rangle & \dot{=} & \left(\begin{array}{c} \cos\frac{\theta}{2}\\ e^{i\phi}\sin\frac{\theta}{2}\end{array}\right).
\end{eqnarray}
The parameterization in terms of angles is useful as this state is the positive eigenstate of spin projection about the axis defined by polar and azimuthal angles $\theta$ and $\phi$, respectively. The Pauli operators take the representations
\begin{eqnarray}
\hat{\sigma}^{x} & \dot{=} & \left(\begin{array}{cc} 0 & 1 \\ 1 & 0 \end{array}\right),\nonumber\\
\hat{\sigma}^{y} & \dot{=} & \left(\begin{array}{cc} 0 & -i \\ i & 0 \end{array}\right),\nonumber\\
\hat{\sigma}^{z} & \dot{=} & \left(\begin{array}{cc} 1 & 0 \\ 0 & -1 \end{array}\right).\label{eq:pauli}
\end{eqnarray}
To eliminate the presence of the imaginary unit, one might try to use a higher-dimensional representation in which the real and imaginary parts of $\alpha$, $\beta$ are separate components. That is, one could rewrite two complex components as four real quantities,
\begin{eqnarray}
|\psi\rangle_{r}& \dot{=} & \left(\begin{array}{c} \mbox{Re}(\alpha)\\ \mbox{Im}(\alpha)\\  \mbox{Re}(\beta)\\ \mbox{Im}(\beta) \end{array}\right).\label{eq:realimag}
\end{eqnarray}
The representation in Eq.~(\ref{eq:realimag}) is somewhat analogous to separating wave functions into real and imaginary parts as discussed above. For any complex $z=x+iy$, one has $iz = -y + ix$. Effectively, the action of $i$ can be represented by a {\it real} matrix
\begin{eqnarray}
{\bf i}_{r} & \dot{=} & \left(\begin{array}{cccc} 0 & -1 & 0 & 0 \\ 1 & 0 & 0 & 0\\ 0 & 0 & 0 & -1\\ 0 & 0 & 1 & 0 \end{array}\right).\label{eq:ir}
\end{eqnarray}
The Pauli operators can also be represented as
\begin{eqnarray}
\hat{\sigma}_{r}^{x} & \dot{=} & \left(\begin{array}{cccc} 0 & 0 & 1 & 0 \\ 0 & 0 & 0 & 1\\ 1 & 0 & 0 & 0 \\ 0 & 1 & 0 & 0 \end{array}\right),\nonumber\\
\hat{\sigma}_{r}^{y} & \dot{=} & \left(\begin{array}{cccc} 0 & 0 & 0 & 1 \\ 0 & 0 & -1& 0 \\ 0 & -1 & 0 & 0 \\ 1 & 0 & 0 & 0\end{array}\right)\nonumber\\
\hat{\sigma}_{r}^{z} & \dot{=} & \left(\begin{array}{cccc} 1 & 0 &0 & 0 \\ 0 & 1 & 0 & 0  \\ 0 & 0 & -1 &0 \\ 0 & 0 & 0 & -1 \end{array}\right).\label{eq:pauli2}
\end{eqnarray}
One can show that these operators, together with the real-valued matrix representation of the imaginary unit, satisfy the usual spin algebra. For example, $\left[\hat{\sigma}^{\alpha}_{r}\right]^{2} = \hat{I},$ and
\begin{eqnarray}
\left[\hat{\sigma}_{r}^{x},\hat{\sigma}_{r}^{y}\right] & = & 2{\bf i}_{r}\hat{\sigma}_{r}^{z},\;\;\;\;
\left[\hat{\sigma}_{r}^{y},\hat{\sigma}_{r}^{z}\right] = 2{\bf i}_{r}\hat{\sigma}_{r}^{x},\;\;\;\;
\left[\hat{\sigma}_{r}^{z},\hat{\sigma}_{r}^{x}\right] = 2{\bf i}_{r}\hat{\sigma}_{r}^{y}.\nonumber\\\label{eq:spincommute}
\end{eqnarray}
Furthermore, it is a simple matter to show that ${\bf i}_{r}$ actually commutes with each of the Pauli operators. Thus, it appears that a perfectly well-defined representation for spin-$\frac{1}{2}$ states is possible without the explicit appearance of $i=\sqrt{-1}$. But potential problems emerge when considering states of more than one spin. Given two pure, single-spin states, $|\psi_{1}\rangle$ and $|\psi_{2}\rangle$, the total state is given by the tensor product $|\psi\rangle = |\psi_{1}\rangle \otimes |\psi_{2}\rangle$. This operation can be represented by the Kronecker product between individual complex vectors. Conveniently, the Kronecker product operation is built into most modern computational environments such as \textsc{Matlab},\cite{kronml} the \textsc{Wolfram} programming language,\cite{kronma} and the \textsc{NumPy} Python library.\cite{kronnp} The explicit representation for the Kronecker product of two, two-dimensional vectors is
\begin{eqnarray}
|\psi\rangle & \dot{=} & \left(\begin{array}{c} \alpha_{1}\\ \beta_{1}\end{array}\right)\otimes\left(\begin{array}{c} \alpha_{2}\\ \beta_{2}\end{array}\right)\nonumber\\
& = & \left( \begin{array}{cc} \alpha_{1}\alpha_{2}\\ \alpha_{1}\beta_{2}\\ \beta_{1}\alpha_{2}\\ \beta_{1}\beta_{2}\end{array}\right).
\end{eqnarray}
Thus, a two-spin state is represented by four complex values, or eight real values. Employing the same Kronecker product to the four-dimensional, real representation, one obtains a 16-component vector representing the composite state. The real vector contains {\it twice} the number of (real) values needed to describe the state using the complex representation. This discrepancy suggests that either there exists significant redundancy in this real-valued representation or the two representations are not entirely equivalent. That is, there {\it might} be some experiment for which the two representations could conceivably give different predictions. One also might attempt to modify the action of the Kronecker product in the real representation to maintain agreement with the complex representation, but we do not pursue that approach in this work.
\subsection{Testing a real-valued quantum theory}
In this section we discuss a series of Bell tests involving the Clauser-Horne-Shimony-Holt (CHSH) inequality\cite{CHSH} to test the predictions of the real-valued description of quantum theory against those of the traditional, complex-valued description. Recently, a particularly clear and self-contained description of the CHSH inequality appropriate for introductory students has was published in {\it The Physics Teacher}.~\cite{BrodyTPT} In addition to deriving the inequality from first principles, the authors show how to test it on a quantum computer. The CHSH inequality is formulated by considering two observers, Alice and Charlie,\footnote{The traditional choice for the second observer is Bob, but we employ Charlie for this role to make the notation consistent with the scenario in the next section in which Bob plays a different role.} who share an entangled spin state: $|\Psi^{-}\rangle = (\left|\uparrow\downarrow\right\rangle - \left|\downarrow\uparrow\right\rangle)/\sqrt{2}$. One spin is sent to Alice and the other to Charlie. The separation between Alice and Charlie is typically assumed to be arbitrarily large to forbid any possible communication of information during the experiment. Alice randomly chooses one of two predetermined observables ($\hat{A}_{1}$ or $\hat{A}_{2}$) to measure. Charlie also randomly chooses one of two predetermined observables ($\hat{C}_{1}$ or $\hat{C}_{2}$) to measure which satisfy nontrivial commutation relations with Alice's observables. Each of the measured observables corresponds to a particular component of spin. Conventionally, the factor of $\frac{\hbar}{2}$ is dropped so that the outcome of each individual measurement is $\pm1$. After many repetitions, Alice and Charlie may compare results. 

It is well known that operators corresponding to orthogonal spin components do not commute and cannot be observed simultaneously. In a Stern-Gerlach experiment, one might begin with the state $\left|\uparrow\right\rangle$. Any subsequent measurement of $\hat{\sigma}^{z}$ returns the value $+1$. However, upon measuring $\hat{\sigma}^{x}$, one obtains $\pm 1$ with equal probability. This statement can only be verified by repeating the experiment (beginning with $\left|\uparrow\right\rangle$) many times. 

Local realism posits that this apparent randomness reflects our own ignorance of the details of some ``hidden variables.'' It is helpful to think of these hidden variables as internal degrees of freedom which are hidden from us but that fully determine the outcome of the subsequent measurement of $\hat{\sigma}^{x}$. Local realism requires that the outcome of any measurement is entirely deterministic and explains the apparent randomness as being due to our ignorance of hidden details. 

To state the CHSH inequality, let us consider the following operator
\begin{eqnarray}
\hat{\beta} = \hat{A}_{1}\otimes \hat{C}_{1} + \hat{A}_{1}\otimes\hat{C}_{2} +\hat{A}_{2}\otimes\hat{C}_{1}-\hat{A}_{2}\otimes\hat{C}_{2}.\label{eq:chshop}
\end{eqnarray}
Assuming local realism results~\cite{BrodyTPT} in the restriction $|\langle \Psi^{-} | \hat{\beta} |\Psi^{-}\rangle | \leq2$, which is known as the CHSH inequality. Letting $\hat{A}_{1} = \hat{\sigma}^{z}$, $\hat{A}_{2} = \hat{\sigma}^{x}$, $\hat{C}_{1} = -(\hat{\sigma}^{z}+\hat{\sigma}^{x})/\sqrt{2}$, $\hat{C}_{2} = (\hat{\sigma}^{x}-\hat{\sigma}^{z})/\sqrt{2}$, the CHSH operator simplifies to
\begin{eqnarray}
\langle \hat{\beta}\rangle & = & -\sqrt{2}\langle \left[\hat{\sigma}^{x}\otimes\hat{\sigma}^{x} + \hat{\sigma}^{z}\otimes\hat{\sigma}^{z}\right]\rangle.
\end{eqnarray}
The state $|\Psi^{-}\rangle$ defined above is actually a spinless state, resulting in perfect anticorrelation between any two like spin components. That is, $\langle\Psi^{-}|\hat{\sigma}^{x}\otimes\hat{\sigma}^{x}|\Psi^{-}\rangle = -1$, as may be shown by direct calculation. This anticorrelation leads to $\langle \Psi^{-}|\hat{\beta}|\Psi^{-}\rangle = 2\sqrt{2}\approx 2.828$, in direct violation of the CHSH inequality. There exist upper limits known as Tsirelson bounds~\cite{Tsirelson} for quantum correlations such as those on the right-hand side of Eq.~(\ref{eq:chshop}). It can be shown that $\langle\hat{\beta}\rangle \leq 2\sqrt{2}$ so that the state $|\Psi^{-}\rangle$ actually achieves maximal possible violation of the CHSH inequality.

Notice that the argument above does not make use of any complex numbers in the representations of the state or the operators. One may also use Eqs.~(\ref{eq:realimag})--(\ref{eq:pauli2}) to show that the real representation described above yields $\langle \Psi^{-}_{r} |\hat{\beta}_{r}|\Psi^{-}_{r}\rangle = 2\sqrt{2}$. A Jupyter notebook to perform this computation directly in terms of real representations of $\hat{A}_{j}$, $\hat{B}_{j}$, and $|\Psi^{-}\rangle$ is included in the Supplementary Material. Thus, there is no tension between the two descriptions even in this two-spin setting. 

To explore a scenario in which complex coefficients play a crucial role, we can consider the ``triple'' CHSH operator\cite{AcinPRA,BowlesPRA} 
\begin{eqnarray}
\hat{\beta}^{(3)} & = & \hat{A}_{1}\otimes\hat{C}_{1} + \hat{A}_{1}\otimes\hat{C}_{2} + \hat{A}_{2}\otimes\hat{C}_{1}-\hat{A}_{2}\otimes\hat{C}_{2}\nonumber\\
& + &  \hat{A}_{1}\otimes\hat{C}_{3} + \hat{A}_{1}\otimes\hat{C}_{4} -\hat{A}_{3}\otimes\hat{C}_{3}+\hat{A}_{3}\otimes\hat{C}_{4}\nonumber\\
& + &  \hat{A}_{2}\otimes\hat{C}_{5} + \hat{A}_{2}\otimes\hat{C}_{6} -\hat{A}_{3}\otimes\hat{C}_{5}+\hat{A}_{3}\otimes\hat{C}_{6}.\label{eq:chsh3}
\end{eqnarray}
 Eq.~(\ref{eq:chsh3})  The operators for Alice and Charlie are given by 
\begin{eqnarray}
\hat{A}_{1} & = & \hat{\sigma}^{z},\;\;\;\;\;\; \hat{A}_{2}= \hat{\sigma}^{x},\;\;\;\;\;\;\hat{A}_{3} = \hat{\sigma}^{y},\label{eq:alice}
\end{eqnarray}

\begin{eqnarray}
\hat{C}_{1} & = & \frac{\hat{\sigma}^{z}+\hat{\sigma}^{x}}{\sqrt{2}},\;\;\;\;\;\; \hat{C}_{2}\;= \;\frac{\hat{\sigma}^{z}-\hat{\sigma}^{x}}{\sqrt{2}},\;\;\;\;\;\;\hat{C}_{3}\; =\; \frac{\hat{\sigma}^{z}+\hat{\sigma}^{y}}{\sqrt{2}},\nonumber\\
\hat{C}_{4} & = & \frac{\hat{\sigma}^{z}-\hat{\sigma}^{y}}{\sqrt{2}},\;\;\;\;\;\; \hat{C}_{5} \;=\; \frac{\hat{\sigma}^{x}+\hat{\sigma}^{y}}{\sqrt{2}},\;\;\;\;\;\;\hat{C}_{6}\; =\; \frac{\hat{\sigma}^{x}-\hat{\sigma}^{y}}{\sqrt{2}}.\nonumber\\ \label{eq:charlie}
\end{eqnarray}
Since Eq.~(\ref{eq:chsh3}) is a sum of three CHSH operators,\cite{BowlesPRA} local realism constrains $|\langle \hat{\beta}^{(3)}\rangle |\leq 3\times2=6$. The operator $\hat{\beta}^{(3)}$ appears rather complicated and asymmetric, involving six operators for Charlie but only three operators for Alice. However, Eq.~(\ref{eq:chsh3}) can be simplified considerably in terms of the Pauli operators to
\begin{eqnarray}
\hat{\beta}^{(3)} & = & 2\sqrt{2}\left[\hat{\sigma}^{x}\otimes\hat{\sigma}^{x} - \hat{\sigma}^{y}\otimes\hat{\sigma}^{y} +  \hat{\sigma}^{z}\otimes\hat{\sigma}^{z}\right].\label{eq:chshop3}
\end{eqnarray}
Maximum violation $\langle \hat{\beta}^{(3)}\rangle \leq 6$ is achieved in traditional quantum theory by $|\Phi^{+}\rangle = (\left|\uparrow\uparrow\right\rangle + \left|\downarrow\downarrow\right\rangle)/\sqrt{2}$.\cite{BowlesPRA} An explicit calculation using \textsc{NumPy} to evaluate expectation value of Eq.~(\ref{eq:chsh3}) is presented in the Supplementary Material. It is also straightforward to calculate the spin expectation values in Eq.~(\ref{eq:chshop3}) using the well-known action of the spin-$\frac{1}{2}$ operators on the states $|\uparrow\rangle$ and $|\downarrow\rangle$. Accordingly, 
\begin{eqnarray}
\langle\Phi^{+}|\hat{\sigma}^{\alpha}\otimes\hat{\sigma}^{\alpha}|\Phi^{+}\rangle & = & \left\{\begin{array}{cc} +1 & (\alpha = x,z)\\ -1 & (\alpha = y)\end{array}\right.
\end{eqnarray}
Thus, we obtain $\langle\Phi^{+}|\hat{\beta}^{(3)}|\Phi^{+}\rangle = 6\sqrt{2}\approx 8.485$ for the complex-valued representation of quantum theory. Using the real-valued representation Eq.~(\ref{eq:pauli2}) leads to identical results for all terms in Eq.~(\ref{eq:chshop3}) {\it except} for
\begin{eqnarray}
\langle\Phi_{r}^{+}|\hat{\sigma}_{r}^{y}\otimes\hat{\sigma}_{r}^{y}|\Phi_{r}^{+}\rangle = 0,
\end{eqnarray}
where $|\Phi_{r}^{+}\rangle = \left(\left|\uparrow\uparrow\right\rangle_{r} + \left|\downarrow\downarrow\right\rangle_{r} \right)/\sqrt{2}$. We employ the shorthand $\left|\uparrow\uparrow\right\rangle_{r} \equiv \left|\uparrow\right\rangle_{r}\otimes\left|\uparrow\right\rangle_{r}$ with $\left|\uparrow\right\rangle_{r}$ defined below (and similarly for $\left|\downarrow\downarrow\right\rangle_{r}$). Strictly speaking, this gives $\langle\Phi_{r}^{+}|\hat{\beta}_{r}|\Phi_{r}^{+}\rangle = 4\sqrt{2}\approx5.657$, failing to violate the triple CHSH inequality. The discrepancy can be traced back to the way in which $\hat{\sigma}^{y}_{r}$ acts. Representing
\begin{eqnarray}
\left|\uparrow\right\rangle_{r} & \dot{=} & \left(\begin{array}{c} 1 \\ 0\\ 0\\ 0\end{array}\right), \;\;\;\;\;\;\;\; \left|\downarrow\right\rangle_{r} \;\dot{=} \; \left(\begin{array}{c} 0 \\ 0\\ 1\\ 0\end{array}\right), 
\end{eqnarray}
the action of $\hat{\sigma}^{y}_{r}$ is
\begin{eqnarray}
\hat{\sigma}_{r}^{y}\left|\uparrow\right\rangle_{r} &\dot{=}& \left(\begin{array}{c} 0 \\ 0 \\ 0 \\ 1\end{array}\right).\label{eq:idown}
\end{eqnarray} 
Consulting Eq.~(\ref{eq:realimag}), we see that this vector is a representation of $i\left|\downarrow\right\rangle$. Indeed, one may show ${\bf i}_{r}\left|\downarrow\right\rangle_{r}$ is equivalent to the right-hand side of Eq.~(\ref{eq:idown}). The {\it problem} is that while $\left\langle \downarrow \right| i \left|\downarrow\right\rangle = i$, we have $_{r}\left\langle \downarrow\right| {\bf i}_{r} \left|\downarrow\right\rangle_{r} = 0$. 

The states $\left|\uparrow\right\rangle$ and $i\left|\uparrow\right\rangle$ should be physically indistinguishable. But these states have orthogonal vector representations in the real-valued theory. As our modest aim is only to sketch how difficulties can arise with a real-valued formulation of quantum theory, we do not investigate here whether a general prescription exists for ensuring global phase invariance of physical states in this representation. However, it is straightforward to show that a suitably modified initial state can still yield maximum violation of the triple CHSH inequality. Let us define
\begin{eqnarray}
\left|\uparrow'\right\rangle_{r} & \dot{=} & \left(\begin{array}{c} 0 \\ 1\\ 0\\ 0\end{array}\right), \;\;\;\;\;\;\;\; \left|\downarrow'\right\rangle_{r} \;\dot{=} \; \left(\begin{array}{c} 0 \\ 0\\ 0\\ 1\end{array}\right).
\end{eqnarray}
Then taking $|\Phi_{r}^{+'}\rangle = \left(\left|\uparrow'\uparrow'\right\rangle_{r} + \left|\downarrow'\downarrow'\right\rangle_{r} \right)/\sqrt{2}$, we let $\left|\Phi_{r}\right\rangle = (|\Phi_{r}^{+}\rangle + |\Phi_{r}^{+'}\rangle)/\sqrt{2}$. A straightforward calculation (see Supplementary Material for details) yields $\langle \Phi_{r}|\hat{\beta}_{r}^{(3)}|\Phi_{r}\rangle = 6\sqrt{2}$, demonstrating that the real-valued representation is technically capable of predicting the same CHSH-violating correlations as the complex-valued representation for this scenario. 

In the next section, we discuss a generalized type of Bell test which has been proven rigorously to yield an upper bound for correlations produced by real-valued quantum theory which can be tested experimentally. 

\section{Testing real-valued quantum theory on a quantum computer}\label{sec:exp}
A Bell-like test has been developed~\cite{SciAm,Renou2021} in which {\it any} real-valued formulation of quantum theory predicts a different outcome than complex-valued quantum theory. We emphasize that by ``real-valued quantum theory,'' we mean a formulation based on the postulates in Sec.~\ref{sec:real} in which the complex-valued state space is restricted to only involve real coefficients. One type of real-valued construction~\cite{McKaguePRL} which uses auxiliary qubits to encode complex phases within a real Hilbert space is capable of matching all predictions of complex-valued quantum theory. But this construction does not satisfy the tensor-product rule for composite systems. The experiment described in this section cannot invalidate such real-valued representations, and we focus on theories which respect the tensor-product rule.

\begin{figure}
\begin{center}
\includegraphics[width=8.6cm]{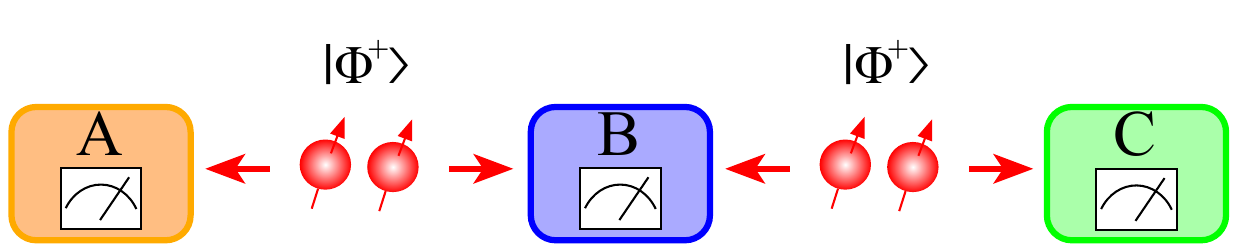}
\caption{Experimental setup: independent sources each create entangled Bell states $|\Phi^{+}\rangle = (|00\rangle + |11\rangle)/\sqrt{2}$. From each state, one qubit is sent to Bob (B) who performs a measurement in the Bell state basis. One of the remaining qubits is sent to Alice (A), and the other is sent to Charlie (C). After Bob's measurement, the residual entanglement is transferred to the qubit pair shared by Alice and Charlie. Alice performs a measurement of $\hat{A}_{\alpha}$, while Charlie performs a measurement of $\hat{C}_{\gamma}$.}
\label{fig:1}
\end{center}
\end{figure}

To align our notation with that of the relevant literature, we adopt the quantum computing convention of mapping spins to the ``computational basis'' states as $\left|\uparrow\right\rangle\rightarrow |0\rangle$, $\left|\downarrow\right\rangle \rightarrow |1\rangle$. The setup for this experiment is depicted in Fig.~\ref{fig:1}. Here, one considers two independent sources, each of which creates the entangled Bell state $|\Phi^{+}\rangle = (|00\rangle + |11\rangle)/\sqrt{2}$. From each of these two entangled states, one qubit is sent to observer Bob. Bob performs a two-qubit measurement in the Bell-state basis, or ``Bell-state measurement,'' from which he concludes that the two-qubit system is in one of the following states:
\begin{eqnarray}
|\Phi^{\pm}\rangle & = & (|00\rangle \pm |11\rangle)/\sqrt{2},\nonumber\\
|\Psi^{\pm}\rangle & = & (|01\rangle \pm |10\rangle)/\sqrt{2}.
\end{eqnarray}
The procedure for Alice and Charlie is almost identical to the triple CHSH experiment described in Sec.~\ref{sec:real}. Alice performs a measurement of $\hat{A}_{\alpha}$ Eq.~(\ref{eq:alice}) on one of the two remaining qubits with $\alpha \in \left\{1,2,3\right\}$ chosen randomly from her three operators. Charlie performs a measurement of $\hat{C}_{\gamma}$ on the remaining qubit with $\gamma\in\left\{1,2,3,4,5,6\right\}$ chosen randomly from his set of six operators. The outcomes of measurements made by Alice ($a = \pm 1$) and Charlie ($c = \pm 1$) depend on the choice of operators ($\alpha\gamma$) and now (due to residual entanglement) on the result of Bob's measurement $\hat{B}$, which maps the states $|\Phi^{\pm}\rangle$, $|\Psi^{\pm}\rangle$ to the binary values $b \equiv b_{1}b_{2}\in\left\{00,01,10,11\right\}$, as described below. Note that $a,c\in \left\{-1,1\right\}$ are each integers whereas $b$ is a bit string. 

Experimentally, Alice, Bob, and Charlie aim to construct a violation of some sort of Bell inequality from these measurements. We define the joint conditional probability $P(abc|\alpha\gamma)$ as the probability that Alice obtains $a$, Bob obtains $b$, and Charlie obtains $c$ given a choice $\alpha\gamma$ for Alice's and Charlie's operators. In terms of the (measurable) quantity $P(abc|\alpha\beta\gamma)$, one may define a particular correlation function $\Gamma$ (analogous to $\langle\hat{\beta}^{(3)}\rangle$) as~\cite{Renou2021}
\begin{eqnarray}
\Gamma = \sum_{abc,\alpha\gamma}w_{abc,\alpha\gamma}P(abc|\alpha\gamma),\label{eq:gamma}
\end{eqnarray}
where $w_{abc,\alpha\gamma} = \pm 1$ are specially-chosen weights,\cite{Renou2021,qmtest1} and $\alpha\gamma \in \left\{ 11, 12,21,22, 13,14,33,34,25,26,35,36\right\}$ give the measurement configuration for Alice and Charlie. For example, $\alpha\gamma = 36$ corresponds to measurements of $\hat{A}_{3}$ in Eq.~(\ref{eq:alice}) and $\hat{C}_{6}$ in Eq.~(\ref{eq:charlie}). The label $abc$ gives the measurement outcomes for all parties so that the sum over $abc$ runs over the entire four-qubit computational basis. To compute $\Gamma$, one defines 

\begin{eqnarray}
\mathcal{T}_{b} & = &  (-1)^{b_{2}}\left(S_{11}^{b} + S_{12}^{b}\right) + (-1)^{b_{1}}\left(S_{21}^{b} - S_{22}^{b}\right)\nonumber\\
& + & (-1)^{b_{2}}\left(S_{13}^{b} + S_{14}^{b}\right) - (-1)^{b_{1}+b_{2}}\left(S_{33}^{b} - S_{34}^{b}\right)\nonumber\\
& +& (-1)^{b_{1}}\left(S_{25}^{b} + S_{26}^{b}\right) - (-1)^{b_{1}+b_{2}}\left(S_{35}^{b} - S_{36}^{b}\right),\label{eq:T}
\end{eqnarray}
where 
\begin{eqnarray}
S_{\alpha\gamma}^{b} = \sum_{a,c=\pm 1}P(abc|\alpha\gamma)ac =\langle \Psi_{0}|\hat{A}_{\alpha}|b\rangle\langle b|\hat{C}_{\gamma}|\Psi_{0}\rangle \nonumber\\\label{eq:S}
\end{eqnarray}
is the expectation value $\langle \Psi_{0}|\hat{A}_{\alpha}\hat{C}_{\gamma}|\Psi_{0}\rangle$ conditioned on the outcome of Bob's measurement. Note that $ac=\pm1$ is simply the product of Alice's and Charlie's measurements, $a$ and $c$, respectively. Finally, the connection between $\Gamma$ and $\mathcal{T}_{b}$ is
\begin{eqnarray}
\Gamma = \sum_{b}\mathcal{T}_{b}.\label{eq:Gam}
\end{eqnarray}
Equations~(\ref{eq:T})--(\ref{eq:Gam}) effectively define the weights $w_{abc,\alpha\gamma}$ in Eq.~(\ref{eq:gamma}), but a simpler interpretation of $\Gamma$ as the expectation value of a particular operator is given below. Though the construction appears somewhat mysterious, it is rather straightforward to implement and provides a clean dichotomy between the real- and complex-valued representations. Defining the initial state $|\Psi_{0}\rangle = |\Phi_{+}\rangle \otimes |\Phi_{+}\rangle$, the standard, complex-valued theory gives $\Gamma = 6\sqrt{2} \approx 8.485$ as shown in the Supplementary Material. The real-valued formulation yields a lower score of $4\sqrt{2}\approx 5.657$ for the same initial state. Unlike with the triple Bell test, combinations of the ``redundant'' states (e.g., involving $\left|\uparrow'\right\rangle$) did not lead to a higher value for $\Gamma$. Readers can explore this calculation for arbitrary initial states by using the Jupyter notebook in the Supplementary Material. 

This discrepancy between the two representations suggests that performing this experiment could confirm whether complex numbers are essential to an accurate description of quantum theory. But the situation is more subtle. We have introduced a {\it particular} real-valued representation in which states in a complex, two-dimensional Hilbert space are mapped to states in a four-dimensional, real space. Other, higher-dimensional representations could be constructed. A priori, it is possible that some representation {\it might} predict the same results as complex-valued quantum theory. 

Remarkably, it has been shown that an upper bound on $\Gamma$ exists for a real representation of {\it any} dimension which obeys the basic postulates (including the tensor-product structure of multiparticle states) summarized in Sec.~\ref{sec:real}. Any such real representation will produce a correlation bounded by $\Gamma \leq \Gamma_{c} \approx 7.6605$.\cite{Renou2021} Interestingly, it is not known how ``tight'' this upper bound is. It should be noted that this correlation bound tacitly assumes additional structure in the theory. In particular, the representation of the tensor product (e.g., the Kronecker product) in the real-valued formulation identical to its action in the complex-valued formulation. While this might seem a natural choice, this similar structure is an underlying assumption of the types of real-valued formulations considered here. 

Experiments using custom-designed superconducting qubits\cite{qmtest1} or photons~\cite{qmtest2} gave convincing evidence that $\Gamma > \Gamma_{c}$. The main goal in the remainder of this section is to show that one can measure $\Gamma$ experimentally using IBM quantum devices. In this way, students with only a minimal working knowledge of IBM quantum computers can perform legitimate experiments that probe fundamental aspects of quantum theory. 

\begin{figure}
\begin{center}
\includegraphics[width=8.6cm]{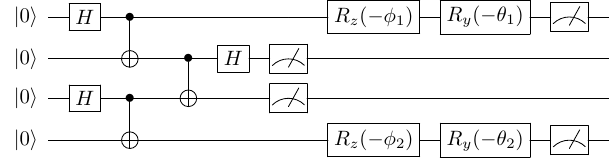}
\caption{Circuit to run on IBM Quantum devices: with each qubit initialized to $|0\rangle$, two pairs of Bell states $|\Phi^{+}\rangle$ are created by the first two sets of Hadamard and CNOT operations. Bob performs a Bell state measurement on a pair constructed from both sets of qubits, while Alice and Charlie make particular measurements on the two remaining qubits. Bob's Bell-state measurement process necessitates the third CNOT and Hadamard set in addition to his computational-basis measurements. Alice and Charlie perform rotations corresponding to their respective operators prior to computational-basis measurements. Twelve copies of this circuit with appropriate rotation angles are required for execution using the \texttt{Sampler} primitive. Only the boxed portion of the circuit (creating the state $|\Psi'\rangle$) is needed when using the \texttt{Estimator} primitive.}
\label{fig:2}
\end{center}
\end{figure}

Figure~\ref{fig:2} depicts a quantum circuit which realizes the experiment shown in Fig.~\ref{fig:1}. Note that the computational basis states are represented by
\begin{eqnarray}
|0\rangle & \dot{=} \left(\begin{array}{c} 1\\ 0\end{array}\right), \;\;\;\;\; |1\rangle \dot{=} \left(\begin{array}{c} 0 \\ 1\end{array}\right).
\end{eqnarray}
The single-qubit Hadamard gate $\hat{H}$ acts according to $\hat{H}|0\rangle = \frac{1}{\sqrt{2}}\left(|0\rangle + |1\rangle\right)$, $\hat{H}|1\rangle = \frac{1}{\sqrt{2}}\left(|0\rangle - |1\rangle\right)$ and takes the matrix representation
\begin{eqnarray}
\hat{H} & \dot{=} \frac{1}{\sqrt{2}}\left(\begin{array}{cc} 1 & 1 \\ 1 & -1\end{array}\right).
\end{eqnarray}
The controlled NOT (CNOT$_{i,j}$) gate flips a target qubit ($j$) only if the control qubit ($i$) is in the state $|1\rangle$, so its matrix representation is
\begin{eqnarray}
\mbox{CNOT}_{i,j} & \dot{=} & \left(\begin{array}{cccc} 1 & 0 & 0 & 0\\ 0 & 1 & 0 & 0 \\ 0 & 0 & 0 & 1\\ 0 & 0 & 1 & 0\end{array}\right).\label{eq:cnot}
\end{eqnarray}
We caution the reader that while we employ $|q_{0}q_{1}q_{2}q_{3}\rangle$ to label a four-qubit state in the circuit in Fig.~\ref{fig:2}, Qiskit uses the opposite state-labeling convention, $|q_{3}q_{2}q_{1}q_{0}\rangle$. Here $q_{0}$ represents the state of the top qubit in the circuit. The convention used in this article better aligns with state-labeling conventions in the physics literature but does require a couple of minor adjustments when implementing expressions in Qiskit, as pointed out in the Supplementary Material. Accordingly, a different matrix representation for CNOT is found in IBM and Qiskit documentation. 

When executing a circuit, IBM hardware initializes each qubit to the $|0\rangle$ state. In order to create the state $|\Phi^{+}\rangle$, one may apply a combination of $\hat{H}\otimes\hat{I}$ and CNOT gates to the state $|00\rangle$ as depicted in Fig.~\ref{fig:2} to obtain
\begin{eqnarray}
\mbox{CNOT}_{i,j}[ \hat{H}\otimes \hat{I} ] |0\rangle_{i}\otimes |0\rangle_{j} & = & \frac{1}{\sqrt{2}}\left(|00\rangle + |11\rangle\right).
\end{eqnarray}
With two copies of $|\Phi^{+}\rangle$, Bob must perform a Bell state measurement on a state consisting of one qubit from each pair. Measurement is only possible in the computational basis on IBM devices, so any observer is only able to determine if a qubit is in the state $|0\rangle$ or the state $|1\rangle$. However, it can be shown that the application of a CNOT gate followed by $\hat{H}$ acting on appropriate qubits converts
\begin{eqnarray}
|\Phi^{+}\rangle & \rightarrow & |00\rangle,\nonumber\\
|\Phi^{-}\rangle & \rightarrow & |10\rangle,\nonumber\\
|\Psi^{+}\rangle & \rightarrow & |01\rangle,\nonumber\\
|\Psi^{-}\rangle & \rightarrow & |11\rangle.
\end{eqnarray}
Thus, Bob's actual result of `00', `01', `10', or `11' allows an unambiguous determination of the two-qubit state in the Bell state basis prior to the gate transformations. Additionally, one notes that $(-1)^{b_{j}}=\pm1$ maps the qubit states $b_{j}=0,1$ to Bob's measurement values $+1,-1$, respectively. Using this observation and Eq.~(\ref{eq:S}), one may recast (see Supplementary Material) the complicated expression for $\Gamma$ as the following expectation value $\Gamma = \langle\Psi'|\hat{\mathcal{O}}|\Psi'\rangle$, where
\begin{eqnarray}
\hat{\mathcal{O}} & = & \hat{\sigma}^{x}\otimes\hat{\sigma}^{z}\otimes \hat{I}\otimes \hat{\sigma}^{x}-\hat{\sigma}^{y}\otimes\hat{\sigma}^{z}\otimes \hat{\sigma}^{z}\otimes \hat{\sigma}^{y}\nonumber\\
& + & \hat{\sigma}^{z}\otimes\hat{I}\otimes\hat{\sigma}^{z}\otimes \hat{\sigma}^{z},\label{eq:op}
\end{eqnarray}
and $|\Psi'\rangle \equiv \mbox{CNOT}_{1,2}\hat{H}_{1}|\Psi_{0}\rangle$ with $\hat{H}_{1}$ indicating that the Hadamard gate acts on $q_{1}$. These two different expressions for $\Gamma$ (Eq.~(\ref{eq:op}) versus Eqs.~(\ref{eq:T})--(\ref{eq:Gam})) motivate two different approaches to measuring $\Gamma$ on IBM hardware. 

As of 2024, all jobs submitted to the quantum hardware must make use of Python objects known as {\it primitives} which produce different types of measurement results. The \texttt{Estimator} primitive produces an estimate of the expectation value of some specified operator(s) (e.g., Eq.~(\ref{eq:op})) for the state defined by the output of a quantum circuit. Alternatively, the \texttt{Sampler} primitive samples the output of a circuit through direct computational-basis measurements, generating a list of counts from which approximate probability distributions (e.g., Eqs.~(\ref{eq:T})--(\ref{eq:Gam})) may be reconstructed. The \texttt{Estimator} primitive is more efficient, allowing automatic error mitigation and requiring virtually no post-processing of the results. However, as this is somewhat of a ``black box'' type of approach, we also sketch how to obtain comparable results using \texttt{Sampler}. 

To minimize errors in the execution of the circuit in Fig.~\ref{fig:2}, it is essential to ensure the physical qubits are connected in such a way that the CNOT gates can be implemented directly. When a small circuit is run, an optimal set of physically connected qubits is selected automatically. Users may consult the job information on their IBM Quantum Dashboard to examine which qubits were used for a job. Additionally, it is possible to manually select which qubits are used (see Supplementary Material). Detailed calibration information for devices used is included in the Supplementary Material.

\subsection{Experiment using \texttt{Estimator}}
In order to use the \texttt{Estimator} primitive, we note that the boxed portion of the circuit in Fig.~\ref{fig:2} creates the state $|\Psi'\rangle$. To represent the operator $\hat{\mathcal{O}}$ in Qiskit, one uses the \texttt{SparsePauliOp} function to define a matrix representation of the operator. One may define an operator for each of Alice's and Charlie's operators in Eqs.~(\ref{eq:alice})--(\ref{eq:charlie}) and simplify the result within Qiskit or define the operator using Eq.~(\ref{eq:op}) directly. A typical term might look like $\hat{\sigma}^{x}\otimes\hat{\sigma}^{z}\otimes \hat{I}\otimes\hat{\sigma}^{y}$. The identity and Pauli matrices are built-in objects, and one constructs an operator corresponding to this combination using
\begin{verbatim}
SparsePauliOp.from_list([("YIZX",1/np.sqrt(2))]).
\end{verbatim}
Note the reversal of order associated with Qiskit's ``opposite'' labeling convention. A Jupyter notebook which contains an implementation of these steps and runs the circuit on both a circuit simulator and quantum hardware is included in the Supplementary Material. Notably, various forms of error mitigation can be applied automatically when using the \texttt{Estimator} primitive by setting the \texttt{resilience\_level} parameter to 0 (no mitigation), 1 (readout-error mitigation), or 2 (several forms of error mitigation). We found the highest level of error mitigation to be somewhat unreliable, producing erroneously high correlations containing significant experimental uncertainties. Consequently, results are described below for resilience level 1 with those for resilience level 0 documented in the Supplementary Material. Each jobs using \texttt{Estimator} on the most recent, 127-qubit devices resulted in a Qiskit runtime usage of roughly 10-20 seconds.

\subsection{Experiment using \texttt{Sampler}}
This approach is closer in spirit to how the published experiments~\cite{qmtest1,qmtest2} were actually performed. By sampling a circuit, the user retains more low-level control over the execution in being able to specify the number of samples (``shots'') and obtaining raw, unmitigated counts. The \texttt{Estimator} primitive is highly efficient, and one might need to employ a large number of shots in order to obtain comparable results, leading to increased execution time with the sampling approach. Also, at the time of writing \texttt{Sampler} does not support resilience levels for user-friendly error mitigation. It is likely that automatic error mitigation options will be introduced in the near future. For this work, we have applied readout-error mitigation manually.

There is significant pedagogical value in performing standard circuit sampling and reconstructing expectation values from the resulting estimate of a probability distribution. Moreover, the sampling approach was essentially the {\it only} way for everyday users to perform such an experiment until IBM introduced primitives in 2023. This method requires significantly more post-processing than the previous approach, but it is more versatile. As more providers make quantum hardware available to the public, it is likely that additional proprietary mechanisms akin to ``primitives'' will follow. Understanding how to compute a fairly complicated quantity using a low-level, sampling method could represent a valuable skill in a rapidly-changing technology sector.

To sample the circuit in Fig.~\ref{fig:2}, we need 12 copies of the circuit with each copy corresponding to one of the 12 measurement configurations for Alice and Charlie needed to measure each of the terms contained Eq.~(\ref{eq:chsh3}). We simulate random sampling of configurations (Alice's and Charlie's random selections) by executing each configuration an equal ($N\sim \mathcal{O}(10^{4})$) number of times. 

To measure Pauli operators $\hat{\sigma}^{j}$ (or combinations thereof), Alice (or Charlie) must rotate the qubit to bring the measurement basis into alignment with the computational basis. That is, for $\hat{\bf n} = \cos\phi\sin\theta \hat{\bf x} + \sin\phi\sin\theta\hat{\bf y} + \cos\theta \hat{\bf z}$, one can measure $\hat{\sigma}^{\hat{\bf n}}$ by first rotating the qubit by $-\phi$ about the $z$-axis (applying $\hat{R}_{z}(-\phi)$) and then rotating by $-\theta$ about the $y$-axis (applying $\hat{R}_{y}(-\theta)$) to bring the relevant measurement into the computational basis. Reference~\onlinecite{Brody} contains a particularly readable discussion of how to measure such quantities on IBM quantum computers in the context of calculating spin correlation functions. For example, $\alpha\gamma = 36$ corresponds to Alice measuring $\hat{A}_{3} = \hat{\sigma}^{y}$ and Charlie measuring $\hat{C}_{6}= (\hat{\sigma}^{x}-\hat{\sigma}^{y})/\sqrt{2}$. To rotate each qubit appropriately so that a computational-basis measurement returns each of these spin projections, Alice must rotate her qubit by $-\phi_{1} = -\frac{\pi}{2}$ about the $z$-axis followed by a rotation of $-\theta_{1} = -\frac{\pi}{2}$ about the $y$ axis. Charlie must rotate his qubit by $-\phi_{2} = \frac{\pi}{4}$ about the $z$-axis and then by $-\theta_{2} = -\frac{\pi}{2}$ about the $y$-axis. A complete list of rotation angles for each choice of $\alpha\gamma$ is given in Table~\ref{tab:angles}.
\begin{center}
\begin{table}[h]
\begin{tabular}{ l | c | c | c | c  }
\hline
\hline
$\alpha\gamma$ & $\phi_{1}$   &  $\theta_{1}$ & $\phi_{2}$ & $\theta_{2}$\\
\hline
11 & 0 & 0 & 0 & $\pi/4$\\
12& 0 & 0 & 0 & $-\pi/4$\\
21& 0 & $\pi/2$ & 0 & $\pi/4$\\
22& 0 & $\pi/2$ & 0 & $-\pi/4$\\
13 & 0 & 0 & $\pi/2$ & $\pi/4$\\
14& 0 & 0 & $-\pi/2$ & $\pi/4$\\
33 & $\pi/2$ & $\pi/2$ & $\pi/2$ & $\pi/4$\\
34 & $\pi/2$ & $\pi/2$ & $-\pi/2$ & $\pi/4$\\
25 & 0 & $\pi/2$ & $\pi/4$ & $\pi/2$\\
26 & 0 & $\pi/2$ & $-\pi/4$ & $\pi/2$\\
35 & $\pi/2$ & $\pi/2$ & $\pi/4$ & $\pi/2$\\
36 & $\pi/2$ & $\pi/2$ & $-\pi/4$ & $\pi/2$\\
\hline
\end{tabular}
\caption{Rotation angles used for all 12 cases of $\alpha\gamma$.}\label{tab:angles}
\end{table}
\end{center}

\section{Results}
\label{sec:res}
\begin{figure}
\begin{center}
\includegraphics[width=8.2cm]{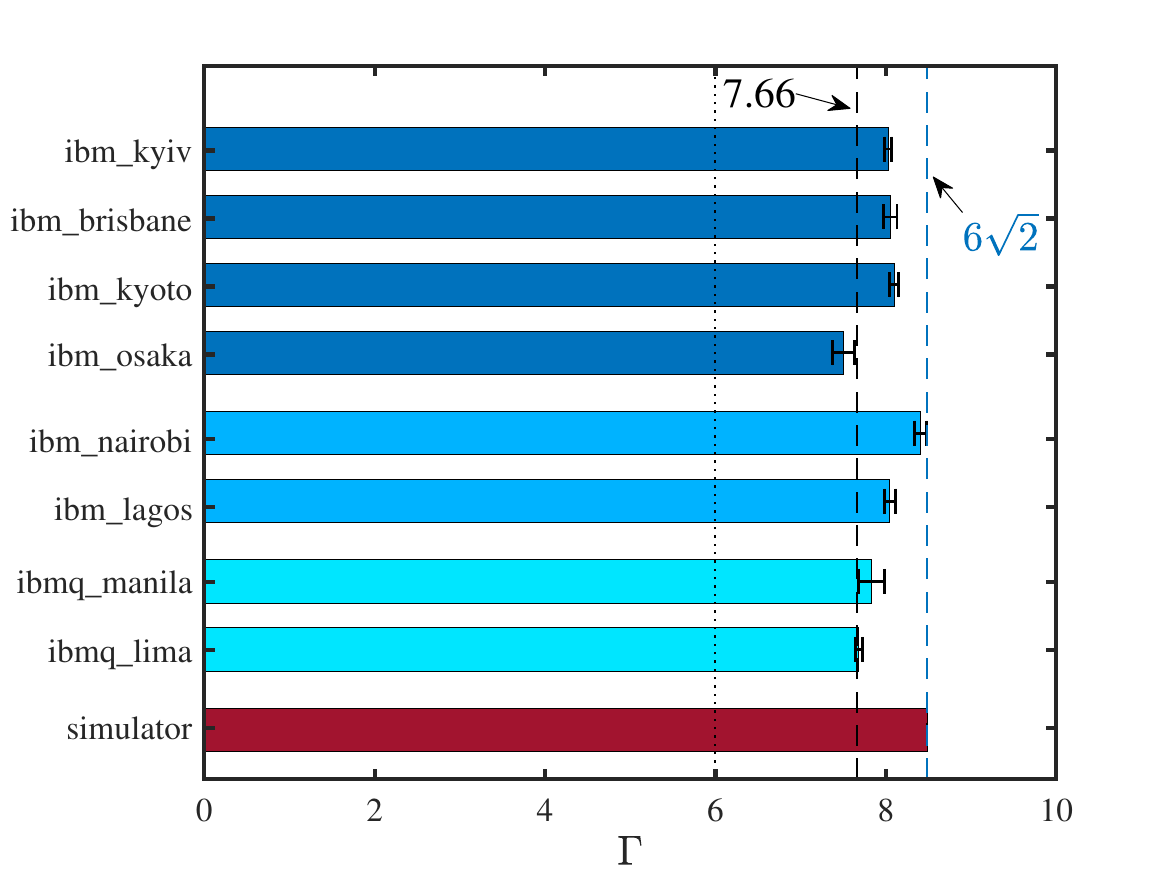}
\caption{Values of $\Gamma$ for the circuit-sampling (QASM) simulator and various IBM devices using \texttt{Sampler} primitive and corresponding results obtained from other hardware using \texttt{Estimator} primitive. Readout-error mitigation has been applied to results for actual devices as described in main text. The lower dotted line denotes the CHSH boundary $\Gamma = 6$,\cite{qmtest1,qmtest2} the lower dashed line at $\Gamma \approx 7.66$ denotes the upper limit for real-valued quantum theory, and the upper dashed line $\Gamma = 6\sqrt{2}$ denotes the value obtained from complex-valued quantum theory. Error bars correspond to $\sigma$ given in Table~\ref{tab:tab1}.}
\label{fig:5}
\end{center}
\end{figure}

Using the \texttt{Estimator} primitive, the expectation value of the operator in Eq.~(\ref{eq:op}) was measured for the state given by the boxed portion of the circuit in Fig.~\ref{fig:2} on the following 127-qubit devices: \texttt{ibm\_osaka}, \texttt{ibm\_kyoto}, \texttt{ibm\_brisbane}, and \texttt{ibm\_kyiv}. This approach yields estimates of $\Gamma$ and the standard error of measurement directly. Results for $\Gamma$ and experimental uncertainty with automatic readout-error mitigation applied (\texttt{resilience\_level} 1) are shown in Table~\ref{tab:tab1} and depicted graphically in Fig.~\ref{fig:5}

\begin{center}
\begin{table}
\begin{tabular}{ l | c | c | c | c | c}
\hline
\hline
 Device &  Method & Trials &  Shots per $\alpha\gamma$ & $\;\Gamma$  & $\sigma$\\
\hline
circuit simulator & sampling & 10 & 32,000 & 8.486 & 0.015\\
\texttt{ibmq\_lima}   & sampling & 9 &20,000  & 7.689 &  0.039\\
\texttt{ibmq\_manila}   & sampling & 21 &20,000 & 7.810 & 0.144\\
\texttt{ibm\_lagos}  & sampling & 44 & 32,000 & 8.065 & 0.060\\
\texttt{ibm\_nairobi}  & sampling & 30 & 32,000 &  8.399 & 0.060 \\
\texttt{ibm\_osaka}  & estimation & 1 & n/a  & 7.500 &  0.128\\
\texttt{ibm\_kyoto}   & estimation & 1 & n/a & 8.097 & 0.057\\
\texttt{ibm\_brisbane} & estimation & 1 & n/a & 8.054 & 0.077\\
\texttt{ibm\_kyiv} &   estimation & 1 & n/a &  8.028 & 0.043 \\
\hline
\end{tabular}
\caption{Measured scores and uncertainty estimates for the circuit-sampling simulator and IBM devices employing basic readout-error mitigation. Sampling shots (per measurement configuration $\alpha\gamma$) are indicated for those devices on which the circuit sampling was employed. Uncertainty corresponds to standard deviation across multiple experiments for results obtained from circuit sampling. For direct estimation, Qiskit provides uncertainty automatically as the standard error. Regardless of device, only four qubits were used in all computations.}\label{tab:tab1}
\end{table}
\end{center}

\begin{figure*}[ht!]
\begin{center}
\includegraphics[totalheight=9.70cm]{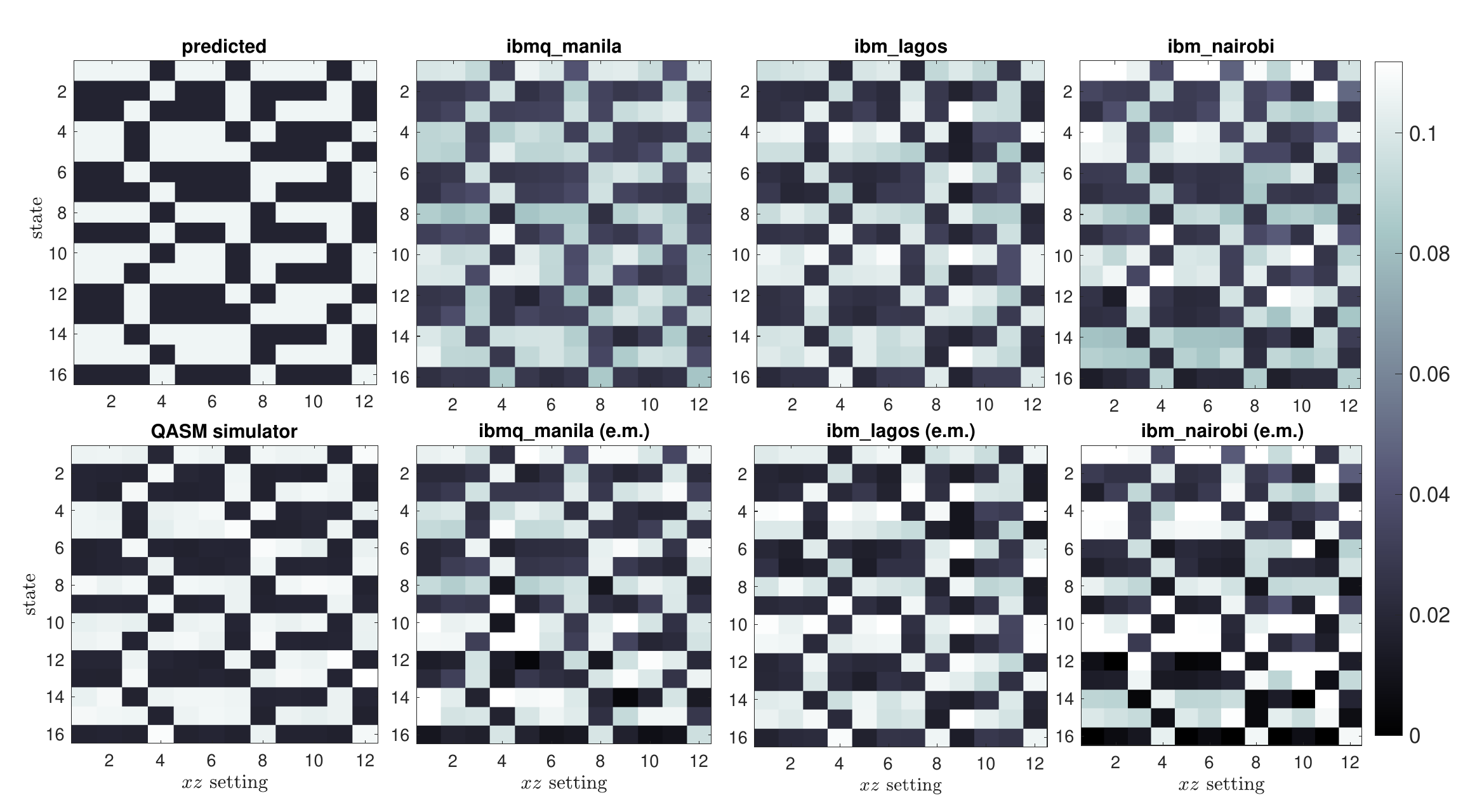}
\caption{Measurements of joint conditional probability, $P(abc|\alpha\gamma)$, where the 16 possible four-qubit states are indexed vertically, while the horizontal axis depicts one of the 12 possible Alice/Charlie measurement combinations $\alpha\gamma$. Results were obtained using \texttt{Sampler} primitive. Theoretical prediction (upper left) follows from standard calculation of the gate operations; QASM simulator refers to IBM's quantum assembly language (QASM) simulator. The label ``e.m.'' on lower plots indicates the use of basic readout-error mitigation. }
\label{fig:prob}
\end{center}
\end{figure*}

Table~\ref{tab:tab1} also shows results from sampling the full circuit in Fig.~\ref{fig:2} using the \texttt{Sampler} primitive on several devices. Each job yielded a list of counts for each of the 16 possible four-qubit states, representing the four single-qubit measurements due to Alice, Bob, and Charlie. The conditional probabilities $P(abc|\alpha\gamma)$ can be reconstructed from this information, and $\Gamma$ can be computed from Eq.~(\ref{eq:gamma}). Samples of reconstructed probability distributions are shown in Fig.~\ref{fig:prob} along with results from the circuit simulator and theoretical predictions. Qiskit's local (QASM) simulator is used to verify that the circuit accurately simulates the experiment in an ideal, noise-free environment with only statistical fluctuations from the simulated counts. While data was collected on smaller devices which have since been retired, the sampling approach required about ten seconds of Qiskit runtime usage for 1024 shots on \texttt{ibm\_brisbane}. When running the same job for 10,000 shots, the Qiskit runtime usage increased to 77 seconds.

With $\mathcal{O}(10^{4})$ shots per choice of $\alpha\gamma$, statistical fluctuations are expected to be small. Noise in the devices leads to errors which can be orders of magnitude larger than statistical fluctuations. We quantify the overall experimental uncertainty roughly by repeating the job at least three times per experiment for three separate experiments over the span of several days, computing the mean $\Gamma$ and standard deviation $\sigma$. These data are summarized in Table~\ref{tab:tab1} and depicted graphically in Fig.~\ref{fig:5}. The resulting uncertainty estimates are comparable to the standard error estimates obtained directly using the \texttt{Estimator} primitive. All data collected, as well as relevant calibration metrics obtained during the measurement events, are tabulated in the Supplementary Material.

Table~\ref{tab:tab1} contains measured values of $\Gamma$ after applying basic, readout-error mitigation. While such error mitigation is incorporated automatically into the results by the \texttt{Estimator} primitive, it was applied manually to the counts obtained with the \texttt{Sampler} primitive. To understand the idea behind readout error mitigation,\cite{Qiskitbook,Geller} note that any job returns a set of ``counts'' as its result. As an example, a two-qubit circuit effectively creates some two-qubit state and measures both qubits. The result of measurement is a list of counts corresponding to the number of times each of the four possible two-qubit basis states ($|00\rangle$, $|01\rangle$, $|10\rangle$, or $|11\rangle$) is observed via computational-basis measurement. In the simplest possible two-qubit circuit, one creates the state $|00\rangle$ and measures both qubits. Supposing this circuit is run $N=1024$ times, the existence of errors means that measurements will not {\it always} correspond to the state $|00\rangle$. For example, one might find $900$ instances of $|00\rangle$, $61$ instances of $|01\rangle$, $54$ instances of $|10\rangle$, and $9$ instances of $|11\rangle$. Repeating this process for all two-qubit basis states, one may build a response matrix ${\bf C}$ for which each of these calibration runs yields a row,
\begin{eqnarray}
{\bf m}_{\mbox{\scriptsize exp}} = {\bf C}{\bf m}_{\mbox{\scriptsize ideal}}.
\end{eqnarray}

Here ${\bf m}_{\mbox{\scriptsize exp}}$ are the obtained counts (in the basis $|00\rangle,\cdots,|11\rangle$), and ${\bf m}_{\mbox{\scriptsize ideal}}$ are the expected results. This matrix ${\bf C}$ can then be inverted and interpreted as an average correction transformation for results obtained from a circuit. Applying ${\bf C}^{-1}$ should approximately correct for systematic measurement errors in the readout process. In this work, we use only this most basic type of error mitigation. A Jupyter notebook performing error mitigation on the sampled results is in the Supplementary Material. 

We now turn to the issue of interpreting results. From Table~\ref{tab:tab1} and Fig.~\ref{fig:5}, it is evident that the noisiest devices (\texttt{ibmq\_lima}, \texttt{ibm\_osaka}, \texttt{ibmq\_manila}), yielded values of $\Gamma$ which fall roughly at the boundary between real-valued and complex-valued quantum theory.\cite{Renou2021} Other devices give values for $\Gamma$ which comfortably exceed the bound of 7.66 when experimental uncertainties are taken into account. Notably, the newest 127-qubit devices give rather consistent results. We can conclude that cloud-based quantum computers provide convincing evidence that real-valued quantum theory is incapable of describing certain observable correlations. 

A potentially confounding aspect of using cloud-based devices to perform experiments via Qiskit is that the user has little control over the calibration and tuning. Strictly speaking, the Qiskit Pulse programming kit~\cite{QiskitPulse} affords users the ability to specify pulse-level timing dynamics in the hardware. Such low-level control could potentially improve the quality of results obtained from these devices. The intent of the present work is to demonstrate that most presently-available devices can deliver reliable results for Bell-test-like experiments even without such low-level control. Only a basic working knowledge of quantum circuits with the simplest form of readout-error mitigation is necessary to obtain the results presented.

IBM devices are calibrated daily, and the relevant parameters are available to users. The calibrations affect the tuning of microwave pulses used to manipulate the qubits, and these daily calibrations attempt to ensure the devices produce reliable and stable results. Data have been collected using the sampling approach across several days during which several calibrations cycles have occurred to sample results at various times between calibrations. Detailed calibration information is collected in the Supplementary Material. The data collected from \texttt{ibm\_lagos} and \texttt{ibm\_nairobi} give results which are consistently above the real/complex threshold by a significant amount ($\sim5\sigma$). 

We note that error mitigation was necessary to obtain values for $\Gamma$ which clearly exceed the upper limit for real-valued quantum theory, $\Gamma_{c} \approx 7.66$. The unmitigated correlations are documented in the Supplementary Material and generally fall below this critical value. However, one of the newest 127-qubit devices, \texttt{ibm\_kyiv}, did yield an uncorrected correlation of $\Gamma = 7.840\pm0.029$ using the \texttt{Estimator} approach. Aside from this anomalously high value, it is interesting to note the consistency of results across generations of devices spanning several years of development.

\section{Discussion}
\label{sec:conc}

We have demonstrated that the freely-available IBM quantum processors are capable of obtaining convincing results for the necessity of complex numbers in the traditional formulation of quantum theory while only requiring the user to arrange a set of basic quantum circuit elements. Most devices do not yield impressive results without readout-error mitigation. But with error mitigation, we have obtained experimental values of $\Gamma$ which approach or (more often) exceed the threshold for the necessity of complex numbers in a traditional formulation of quantum mechanics. While basic error mitigation is required to obtain convincing results, users can now obtain error-mitigated results automatically by selecting the appropriate parameter setting when using the \texttt{Estimator} primitive in Qiskit.

The implications of this experiment on the nature of quantum theory are interesting but rather limited in scope. In essence, the existence of correlations leading to $\Gamma > 7.66$, as observed in this work and in previous experiments,\cite{qmtest1,qmtest2,Wu2022} indicates a failure of real-valued quantum theory. We have followed the terminology of the original proposers~\cite{Renou2021} of this experiment in using the term ``quantum theory'' to denote a framework based on the precise postulates of quantum mechanics stated in Sec.~\ref{sec:real}. We note that the open-access peer-review file which accompanies Ref.~\onlinecite{Renou2021} contains some illuminating dialogue between the authors and the reviewers concerning the limited class of theories which may be falsified by this experiment. Rather than speculate on the viability of more general, real-valued description of quantum phenomena, we present this experiment as an accessible way of constraining certain representations of quantum theory. 

The results obtained from most devices are consistent with previous experimental results.\cite{qmtest1,qmtest2,Wu2022} Recreating or expanding upon our investigations could form the basis for student projects in a course on quantum mechanics. Beyond the experiment presented, the basic framework employed could be used to recreate other experiments involving strongly correlated quantum systems such as demonstrations of multipartite entanglement or nonlocality.\cite{TripartiteMao,TripartiteCao}

\acknowledgements

The authors acknowledge the use of IBM Quantum~\cite{IBMQ} services for this work. The views expressed are those of the author and do not reflect the official policy or position of IBM or the IBM Quantum team. Additionally, the authors acknowledge access to advanced services provided by the IBM Quantum Researchers~\cite{IBMQr} and Educators programs. Lastly, the authors are truly grateful for insights and suggestions received from the anonymous reviewers who evaluated an earlier version of this manuscript.

\section*{Author Declarations}
\subsection*{Conflict of Interest}
The authors have no conflicts to disclose.

\section*{Supplementary Materials}
Supplementary materials including Jupyter notebooks to perform calculations contained in this manuscript can be found at

\href{https://github.com/jllancas/CHSH-quantum-computer}{https://github.com/jllancas/CHSH-quantum-computer}

\end{document}